\documentclass{PoS}
\usepackage{url}

\title{Flasher and muon-based calibration of the GCT telescopes proposed for the Cherenkov Telescope Array.}

\ShortTitle{Calibration of the GCT.}

\author{\speaker{Anthony M. Brown}\\
        Department of Physics, Durham University, DH1 3LE, UK\\
        E-mail: \email{anthony.brown@durham.ac.uk}}

\author{Thomas Armstrong\\
        Department of Physics, Durham University, DH1 3LE, UK\\
	E-mail: \email{thomas.armstrong@durham.ac.uk}}

\author{Paula M. Chadwick\\
        Department of Physics, Durham University, DH1 3LE, UK\\
	E-mail: \email{p.m.chadwick@durham.ac.uk}}

\author{Michael Daniel\\
        Department of Physics, University of Liverpool, Liverpool, L69 7ZE, UK\\
	E-mail: \email{michael.daniel@liverpool.ac.uk}}

\author{Richard White\\
        Max-Planck-Institut fur Kernphysik, D-69029 Heidelberg, Germany\\
	E-mail: \email{richard.white@mpi-hd.mpg.de}}

\author{for the CTA consortium\\
        https://www.cta-observatory.org}

\abstract{The GCT is a dual-mirror Small-Sized-Telescope prototype proposed for the Cherenkov Telescope Array. Calibration of the GCT's camera is primarily achieved with LED-based flasher units capable of producing $\sim4$ ns FWHM pulses of 400 nm light across a large dynamic range, from 0.1 up to 1000 photoelectrons. The flasher units are housed in the four corners of the camera's focal plane and illuminate it via reflection from the secondary mirror. These flasher units are adaptable to allow several calibration scenarios to be accomplished: camera flat-fielding, linearity measurements (up to and past saturation), and gain estimates from both single pe measurements and from the photon statistics at various high illumination levels. In these proceedings, the performance of the GCT flashers is described, together with ongoing simulation work to quantify the efficiency of using muon rings as an end-to-end calibration for the optical throughput of the GCT.}

\FullConference{The 34th International Cosmic Ray Conference,\\
		30 July- 6 August, 2015\\
		The Hague, The Netherlands}

\begin{document}

\section{Introduction}
The Cherenkov Telescope Array (CTA) will be the next generation of ground-based imaging atmospheric cherenkov telescope (IACT) array. Building on the strengths of the current IACTs such as H.E.S.S., MAGIC and VERITAS, CTA is designed to achieve an order of magnitude improvement in sensitivity, with unprecedented energy and angular resolution \cite{cta}. Importantly, CTA will also increase the energy reach of ground-based $\gamma$-ray astronomy, with an energy threshold of $\sim0.02$ TeV and extending to beyond 100 TeV. 

To realise these design goals, CTA will consist of 3 sizes of telescopes. With a diameter of $\sim23$ m, the Large-Sized-Telescope (LST) is designed to observe the low Cherenkov photon flux associated with $E_{\gamma}<0.1$ TeV photon-induced extended air showers (EAS). As such, the LSTs afford CTA with its low energy threshold. The Medium-Sized-Telescopes (MST) will have a diameter of $\sim 12$ m and provide the majority of the improvement in flux sensitivity in the $0.1 \leq E_{\gamma} \leq 1$~TeV energy range. The Small-Sized-Telescopes (SST) will extend the high-energy reach of CTA, predominately observing $\gamma$-rays with energies in excess of 10 TeV. At these energies, the limiting factor is not the number of Cherenkov photons produced in the extended air shower, but rather the number of showers to observe. As such, the SSTs possess the modest mirror diameter of $\sim4$ m, but will be spread over an area greater than $\sim4$ km$^2$, to maximise the effective area of the array.   

The Gamma-ray Cherenkov Telescope (GCT) is a dual-mirror SST (SST-2M) prototype proposed for CTA. A dual-mirror telescope design enables small form factor inexpensive photosensor elements to be used in the camera. Additionally the good point spread function off axis allows a large field of view camera to be built from these photosensors. The Compact High Energy Camera (CHEC \cite{chec}), is the prototype camera for GCT. There are two prototypes for the GCT camera: one based on multi-anode photomultipliers (GCT-M) and one based on silicon photomultipliers (GCT-S). Both of these prototypes will contain 32 modules, totalling 2048 pixels. During event read out, the analogue signal from each pixel is pre-amplified before being passed to the TARGET modules~\cite{target}. In total there are 32 TARGET modules, with each TARGET module possessing 4 custom ASICs that digitise the signal and reads out the full waveform of each pixel, for every event triggering the camera. The use of ASICs in the TARGET module affords the GCT fast data readout, with a large amount of flexibility, allowing for optimisation of the event read-out window, buffering of triggered events and variations in the resolution to which the event waveforms are sampled~\cite{defranco}. 

Throughout the prototyping and commissioning of GCT-S and GCT-M, the performance of observational properties such as pixel linearity, angular response and pixel gain will be \textit{characterised}. However, due to aging effects of the camera components and variations in ambient conditions such as temperature or night sky background (NSB), this initial characterisation will need to be routinely updated during camera operation. When possible, this continual updating of the camera's performance, referred to as \textit{calibration}, should be undertaken with two or more independent and complementary methods to allow for cross-checking of results, and an estimation of systematic uncertainties, as well as to provide a level of redundancy within CTA's calibration scheme. In these proceedings we present two techniques envisaged to be used to calibrate the GCT's camera on nightly timescales: the in-camera LED-based flasher calibration system and images of local muon rings from extended air showers. 

\section{Flasher calibration system}

The primary calibration system for the GCT camera is its LED-based flasher calibration system (FCS). The FCS consists of four identical flasher units housed in the sealed camera body, at the four corners of the camera focal plane (see Figure 1). A Thorlabs ED1-C20 one inch circular top-hat diffuser is mounted in front of the individual flasher units and illumination of the camera focal plane occurs via reflection off the secondary mirror (M2 in Figure 1).

\begin{figure}
 \centering
\includegraphics[width=140mm]{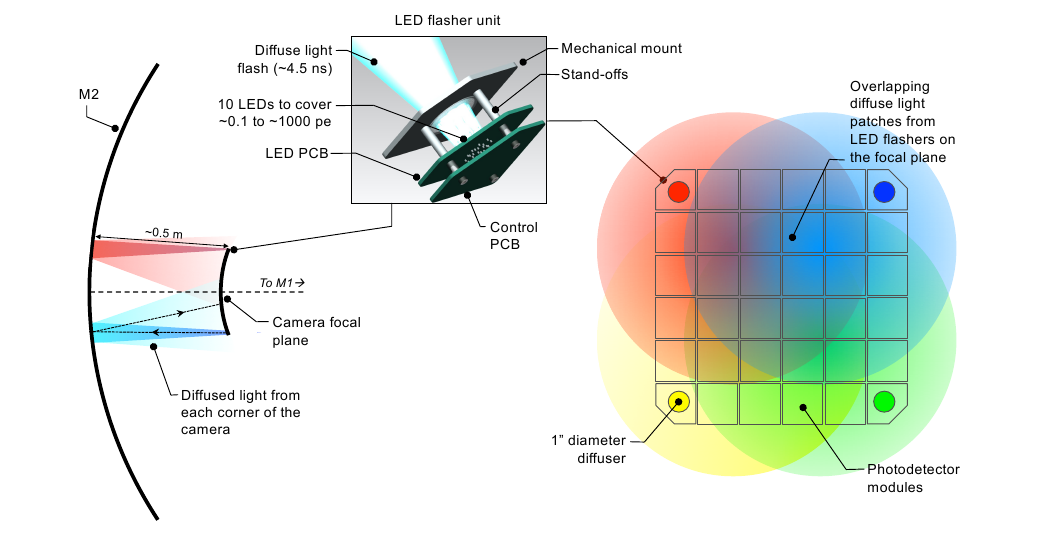}
\caption{\textit{Left}: A schematic of the flasher calibration system's geometry, including the relative position of the secondary mirror (M2) to the focal plane and the use of M2 to illuminate the camera focal plan. \textit{Right}: Schematic to show how the light patterns from the four individual flasher units will be combined, once reflected off M2, to completely cover the camera focal plane. Note that the different colours used here are for illustration purposes only. \textit{Insert}: A CAD model of an individual flasher unit, showing the relative position of the LED PCB, the control PCB, and the mechanical stand-off connected to the diffuser.}
\label{sch}
\end{figure}

Building upon the experience of the VERITAS LED-based flasher unit \cite{veritas}, the GCT's flasher units are a TTL triggered simple gated circuit housing 10 Bivar UV3TZ-400-15 LEDs. Peaking at 400 nm, these LEDs have low self-capacitance, which allows for light pulses with Full-Width-Half-Maximum (FWHM) on the order of $\sim4$ nanoseconds. The flasher units consists of two PCBs: an LED PCB and a control PCB (see Figure 1). The LED PCB houses the 10 LEDs. The control PCB houses a Cypress Semiconductors Programmable System on Chip (PSoC) \footnote{http:$//$www.cypress.com$/$psoc} which takes the external TTL trigger, serial commands and 5V power inputs from GCT camera's peripherals board~\cite{chec,defranco}, sets the required LED pattern and routes the power and TTL trigger through to the LED PCB to produce $\sim4$ ns pulses of UV light.  

\subsection{Dynamic range} 

The dynamic range of the GCT flasher unit's illumination level is designed to be from 0.1 photoelectron (pe) for absolute single-pe calibration, up to $\sim1000$ pe. The GCT FCS has been designed without a filter wheel so as to keep the number of moving parts to a minimum. As such, to achieve the large dynamic range of the GCT flasher units without the use of a filter wheel, the current through individual LEDs is limited by individual resistors ranging from 80 to 140 $\Omega$. 

The distribution of these resistor values has been selected to give a smooth distribution of the flasher unit's log(pe) illumination performance. The measured single-pe spectrum of the sub-pe LED of the GCT flashers, as measured with a SiPM, can be seen in the left plot of Figure 2. The right plot of Figure 2 shows the individual FWHM and pulse area characteristics of the bright LEDs\footnote{Here bright refers to the LEDs producing >10 pe per light pulse.} of the four GCT-M flashers.

The two plots shown in Figure 2 highlight two important performance characteristics of the GCT flashers since the single-pe \textbf{and} large dynamic range capability of the GCT flasher units allows for two independent ways of absolutely calibrating the GCT camera \cite{mirzoyan1997,veritas,mike}. The single-pe capability of the GCT flashers allows us to measure the GCT camera's single-pe response by force triggering the camera in low light conditions (usually in the telescope park position at the start and end of the night's observations). Through the use of photon statistics at different illumination levels, the large dynamic range allows us to calculate the ADCpe ratio by relating the mean and variance of the calibration spectrum to the number of pe (referred to as the F-factor method \cite{mirzoyan1997}). 

\begin{figure*}
\begin{minipage}{150mm}
\centering
\includegraphics[width=.5\textwidth]{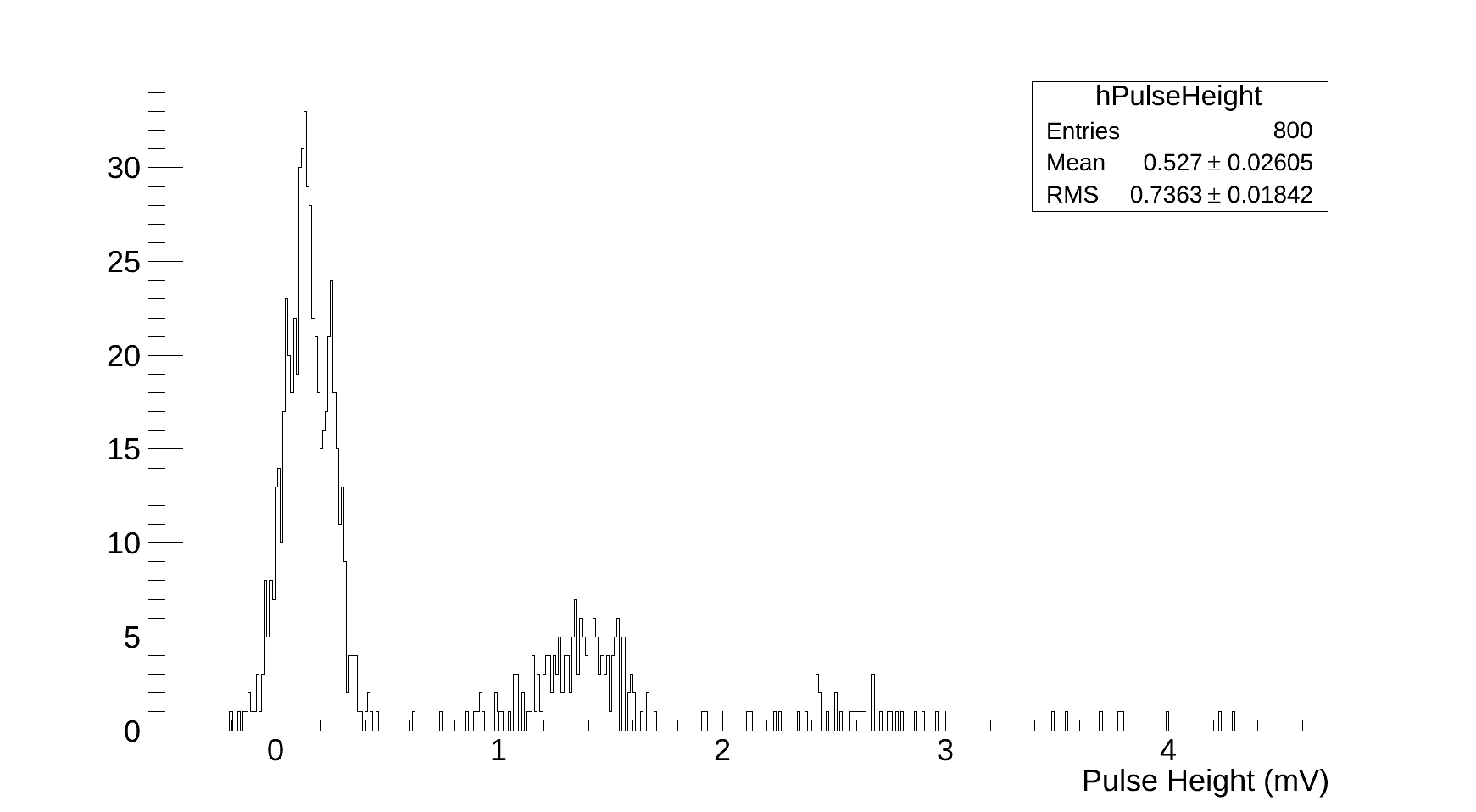}\hfill
\includegraphics[width=.5\textwidth]{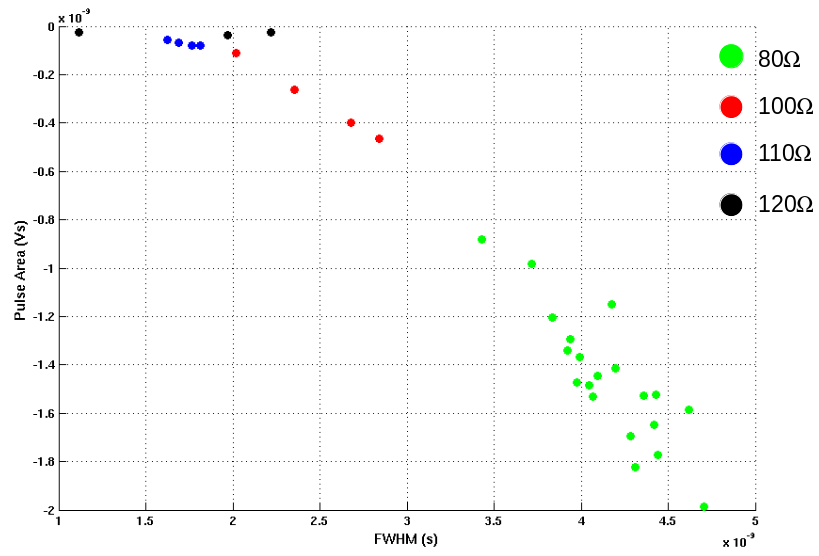}
\caption{\textit{Left}: A pulse area spectra of the single-pe LED of one of the GCT-M flashers, as measured by a SiPM.  \textit{Right}: The pulse area and FWHM values for the individual LEDs of the GCT-M flasher units. Note that the spread in the values for a given resistor is the combination of two known factors: (i) a non-optimised setting of the TTL trigger pulse width for the 4 flasher units for GCT-M, and (ii) variations in LED-to-LED performance. The former factor will be removed with optimsation of the TTL trigger pulses and the latter factor will be reduced by characterising the individual LEDs and optimising the process of populating the LED PCB.}
\end{minipage}
\end{figure*}

\subsection{Ambient Temperature} 

Experience with current IACTs have shown that LEDs are preferable to lasers for use in an on-structure calibration system due to their low-cost, longer lifetime and smaller amount of intrinsic jitter. On the other hand, the absolute light output from LEDs is very temperature dependent. However, given that the GCT's flasher units are located within the sealed camera body which is temperature controlled, variations in the ambient temperature will be greatly mitigated. Nonetheless, the possibility of small variations of the internal camera temperature must still be accounted for. To study the effect of these small temperature variations, the FWHM and light pulse area distributions for the GCT-M flasher units were measured by an MAPM, for the expected operating temperature of the GCT camera. These distributions are shown in Figure 3. As can be seen, in the expected GCT camera operating temperature range of ($18-21$)\ensuremath{^{\circ}}C, there is very little change in the mean of the distribution, with a $\pm 0.02$ns variation in the mean FWHM value and $\pm 0.01$Vns in the mean pulse area value. This implies that the planned temperature stabilisation of the GCT camera is sufficient to minimise systematic error associated with temperature variation of the flasher calibration system. 

\begin{figure*}
\begin{minipage}{150mm}
\centering
\includegraphics[width=.5\textwidth]{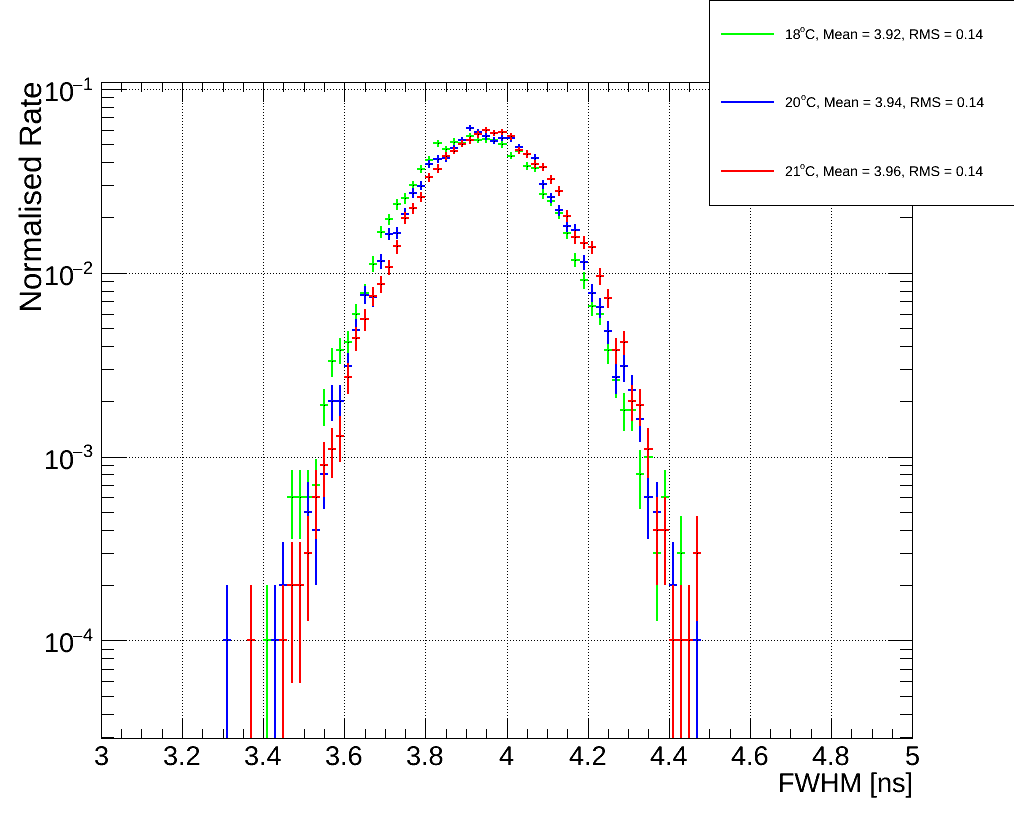}\hfill
\includegraphics[width=.5\textwidth]{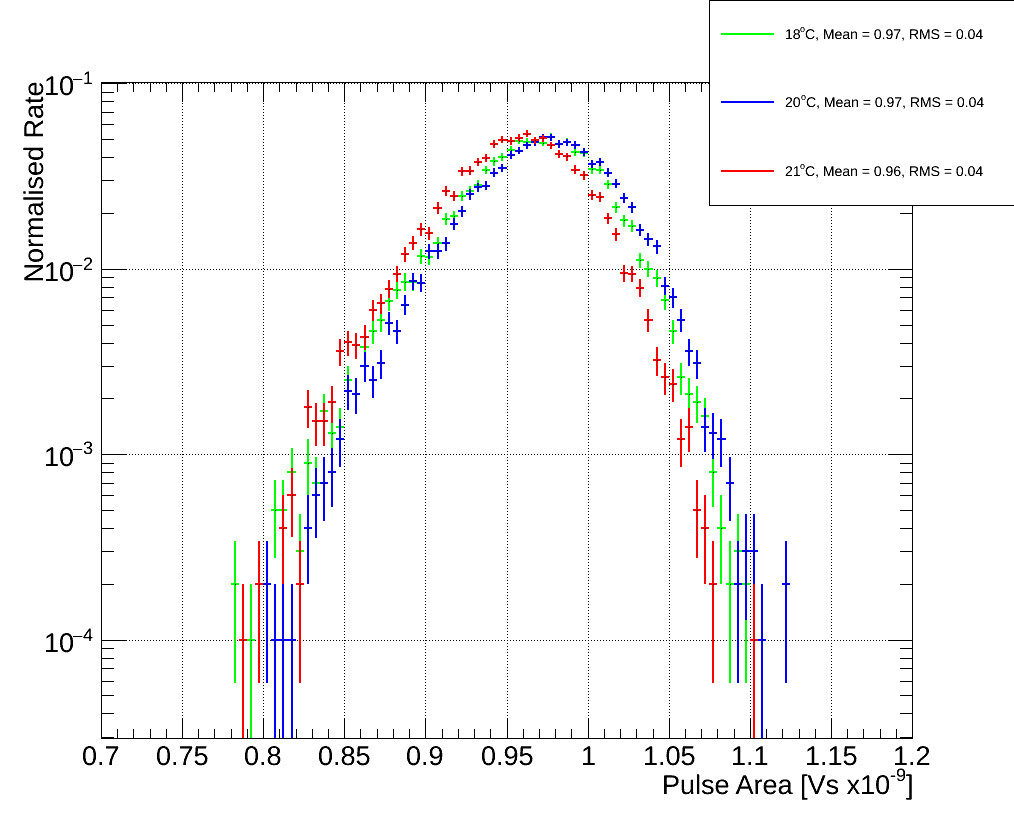}
\caption{\textit{Left}: The FWHM distribution, in nanoseconds, for operating temperatures expected within the GCT-M and GCT-S cameras. \textit{Right}: The pulse area distribution, in Volt-nanoseconds, for operating temperatures expected within the GCT-M and GCT-S cameras. There is no different in the FWHM and pulse area distributions when the ambient temperatures vary by a few degrees relative to the expected maximum operating temperature.}
\end{minipage}
\end{figure*}

\section{Muon ring calibration}

Images of local muon rings have long been used as a calibration method to calibrate, evaluate and monitor the optical throughput of IACTs (eg, see \cite{muons,gaug}). Building upon this experience, CTA will also use muon rings to understand the optical throughput of its telescopes \cite{gaug,simona}. Indeed, given that the amount of Cherenkov light emitted by muons locally\footnote{Here locally refers to the light being emitted within 500m of the telescope.} depends only on the local atmospheric conditions, it can be considered an absolute reference light source for a telescope independent of its type. As such, muon rings provide us with a very powerful tool with which to cross-calibrate the large number of telescopes and the different types of telescope structures envisaged for CTA. 

We undertook a study to investigate the feasibility of using muon rings to calibrate both GCT-M and GCT-S. Properties of the GCT-S and GCT-M prototypes, such as single-pe, photon detection efficiency and quantum efficiency, are discussed in detail in \cite{tom}. With the \textsc{corsika} software package (version 6.99 \cite{cors}), $10^6$ muon ($\mu^-$) events were simulated in a view cone of ($0-4.7$)\ensuremath{^{\circ}} and an impact parameter varying from 0 to 2.2 m (which is the radius of GCT's primary mirror). The muon spectrum was described with a power-law spectrum possessing a spectral index of $\Gamma=-2$, in the energy range from 4 GeV to 1 TeV and the primaries were simulated to originate from an altitude of $\sim2$ km above sea level. The \textsc{sim\_telarray} software package \cite{telarr} was used to simulate the telescope optics and determine when muon events satisfied the first-level trigger level for GCT-M and GCT-S respectively, which were then saved as triggered events. For illustrative purposes, an example of a muon ring image triggering GCT-S can be seen in Figure 4.   

\begin{figure}
\centering
\includegraphics[width=70mm]{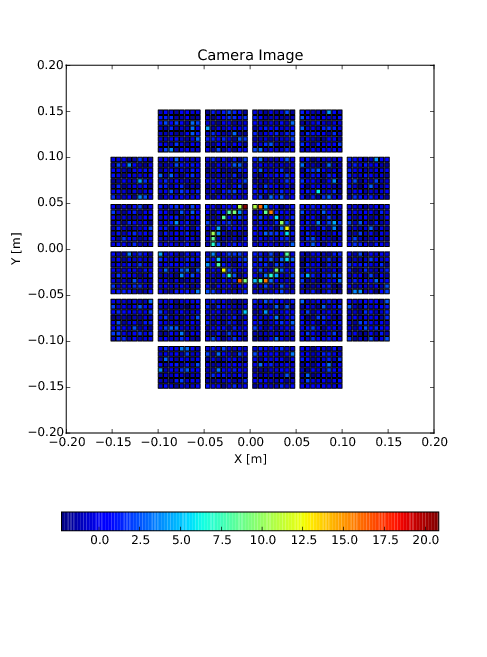}
\caption{An image of a muon ring event, as seen with GCT-S. The colour scale denotes the number of pe detected in a given pixel. The relative x-y position of the pixels and the gaps between the 32 modules, are not representative of the actual GCT focal plane.}
\label{cameraimage}
\end{figure}

Bespoke software was used to conduct a full ring construction analysis of the individual triggered events. In particular, this bespoke software read out the position and amplitude of the triggered pixels, for each triggered muon event, and set the remaining pixels to zero. The resultant images were then modelled with a simple spatial ring model and the errors of the fit were minimised with a Nelder-Mead algorithm. The best-fit ring model was then used to deduce parameters about the muon image such as position within the camera field of view, size of the ring, goodness of the ring model fit, ring completeness and total number of pe within the image. An example of the ring radius resolution of this fitting routine can be seen in the left plot of Figure 5.


\begin{figure*}
\begin{minipage}{150mm}
\centering
\includegraphics[width=.5\textwidth]{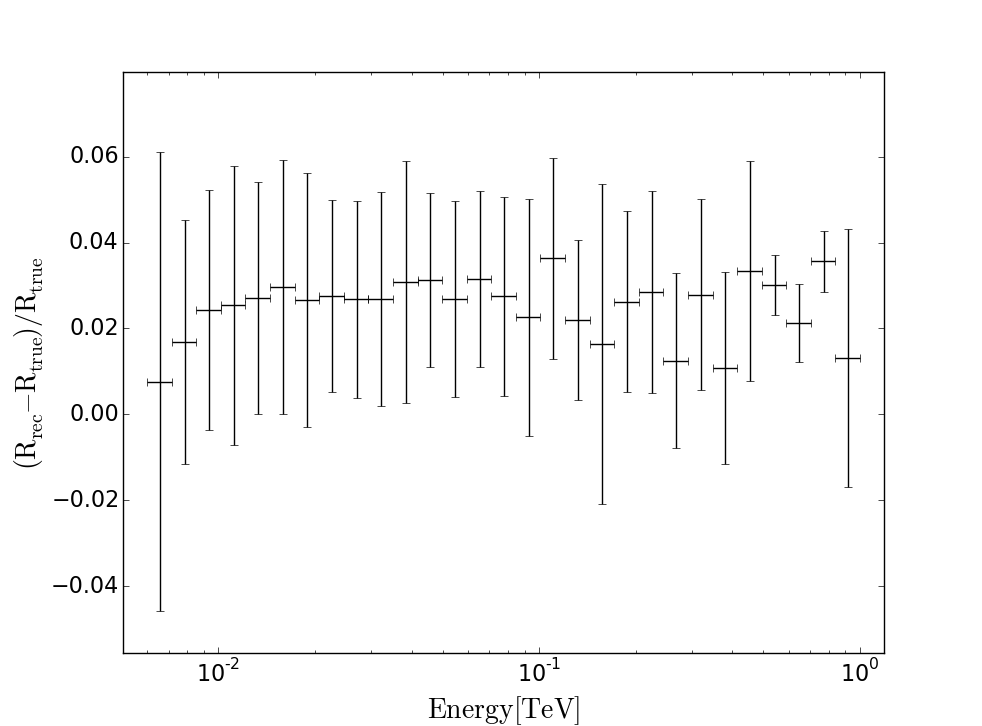}\hfill
\includegraphics[width=.5\textwidth]{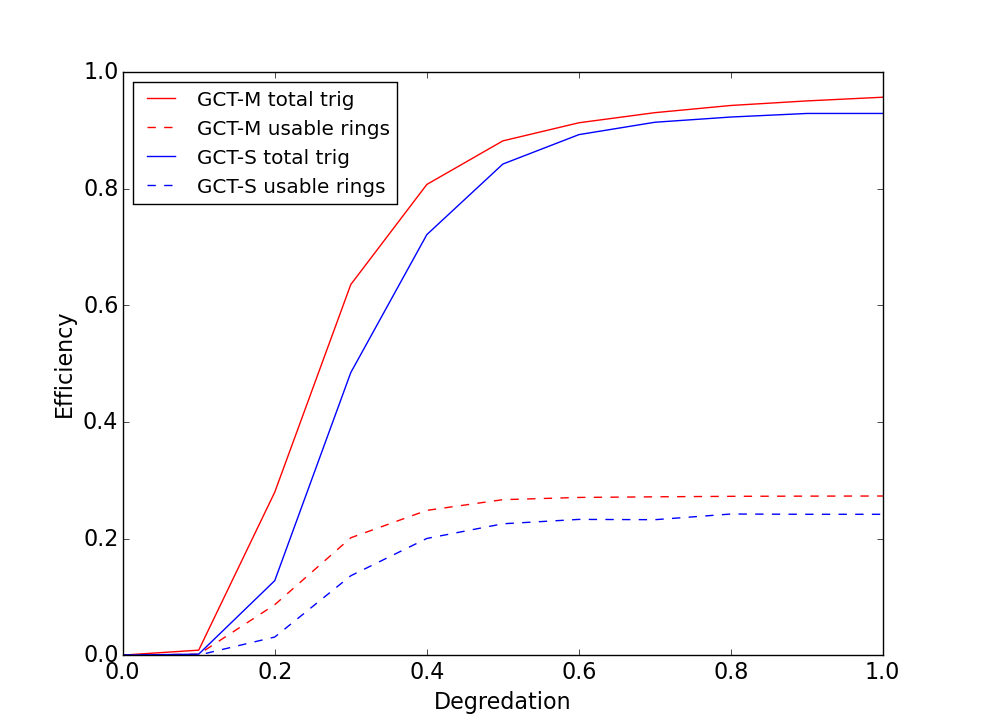}
\caption{\textit{Left}: The resolution of the muon ring radius as a function of muon energy, for GCT-M. The bias on the reconstructed ring radius is at the level of $\pm3$\%, independent of the muon energy. \textit{Right}: The absolute muon trigger efficiency as a function of degradation of the optical efficiency for both GCT-M and GCT-S. The degradation describes the aging of the total optical throughput of the telescope structure, including mirrors. The solid lines denote the trigger efficiency for all events triggering the camera, while the dashed lines indicate the trigger efficiency after cuts have been applied. The dashed lines indicate that the trigger efficiency for both GCT-M and GCT-S is stable down to an optical degradation of 40\%. }
\end{minipage}
\end{figure*}

One important criterion to consideration for the feasibility of using muon rings is how efficiently muon events trigger the GCT as the optical throughput of the telescope ages (primarily due to mirror aging). To investigate this, the optical throughput for the GCT was varied with \textsc{sim\_telarray}, ranging from no degradation (1.0), to complete optical degradation (0.0). For each optical degradation, the total number of triggered events was determined for both GCT-M and GCT-S. The distributions are described by the solid lines in  the right plot of Figure 5, where we see that, below an optical degradation of 60\%, there is a strong dependency of the muon trigger efficiency on the optical throughput of the telescope system. This dependency implies that a small change in the optical degradation results in a large change in the muon trigger rate. However, above 65\% optical throughput there is a flat distribution for both GCT-M and GCT-S\footnote{CTA requires that the minimum average specular reflectivity of an SST reflector, in the 300-550 nm wavelength range, at all times during operation must be $>65$\%.}. This flat feature is important since it implies that the mirrors can lose up to 40\% of their reflectivity, with only a modest change in the muon trigger efficiency. 

In an attempt to flatten the muon efficiency curve further, the following cuts were applied to best-fit ring model parameters of the total triggered events: 

\begin{itemize}
 \item \textbf{Edge Cut}: Any ring that was within $0.3$\ensuremath{^{\circ}} of the edge of the camera were rejected.
 \item \textbf{Amplitude Cut}: Total summed pe of triggered pixels $>40$ pe.
 \item \textbf{Ring completeness}: Only events with $>90$\% pixel coverage around the ring were kept.
\end{itemize}
 
The dashed lines in Figure 5 shows the absolute muon trigger efficiency, as a function of optical degradation of the GCT mirrors, for both GCT-M and GCT-S after these cuts are applied. As can be seen, while these cuts reduce GCT's efficiency of triggering on muons, the dependency of the efficiency on the optical degradation is greatly reduced. Furthermore, above 40\% optical throughput, there is little change in the muon trigger efficiency. While further work is needed to optimise these cuts, the flatness of dashed lines indicates that calibrating the GCT with muon rings is feasible.

\section{Conclusion}

After being characterised in the lab, variations in ambient operating conditions combined with ageing of the camera means that regular calibration of the CTA's cameras is required. While some of this calibration will be done on yearly timescales, other camera calibration aspects such as the absolute calibration of converting observed charge to the number of pe, often referred to as the ADCpe ratio, will be done on shorter timescales. GCT may achieve this through two techniques. 

The primary calibration system for the GCT camera is its LED-based flasher calibration system. GCT's custom built flasher units have both single-pe and large dynamic range capabilities, with $\sim4$ ns FWHM. The single-pe capability allows to directly measure the camera's ADCpe ratio. The large dynamic range allows us to use another method to provide an absolute calibration of the GCT camera in the form of the F-factor method. Finally we note that the GCT's camera response to local muon events, and in particular, comparing this response to detailed simulations, affords us with another method with which to absolutely calibrate the GCT camera.

\textsc{\textbf{Acknowledgements}}: We acknowledge the financial support of the UK's STFC (grant ST/K501979/1) and Durham University. We gratefully acknowledge support from the agencies and organizations listed under Funding Agencies at  \url{http://www.cta-observatory.org}.


\begin{thebibliography}{99}
\bibitem{cta} CTA Consortium, 2013, `\textit{Introducing the CTA concept}', Astroparticle Physics, 43, 3 
\bibitem{chec} Daniel, M.K., et al. 2013, `\textit{A Compact High Energy Camera for the Cherenkov Telescope Array}', Proc. of 33rd ICRC, preprint (arxiv:1307.2807)
\bibitem{target} Bechtol, K., et al. 2012, `\textit{TARGET: A multi-channel digitizer chip for very-high-energy gamma-ray telescopes}', APh, 36, 156
\bibitem{defranco} de Franco, A., et al. 2015, `\textit{The first GCT camera for the Cherenkov Telescope Array}', these proceedings
\bibitem{veritas} Hanna, D., et al. 2010, `\textit{An LED-based flasher system for VERITAS}', NIM A, 612, 278
\bibitem{mirzoyan1997} Mirzoyan, 1997, `\textit{On the calibration accuracy of light sensors in atmospheric Cherenkov fluorescence and neutrino experiment}' Proc. of the 25th ICRC, 265
\bibitem{mike} Daniel, M.K., et al. 2015, `\textit{The camera calibration strategy of the Cherenkov Telescope Array}', these proceedings
\bibitem{muons} Vacanti, G., et al. 1994, `\textit{Muon ring images with an atmospheric Cerenkov telescope}', APh, 2, 1
\bibitem{gaug} Gaug, M., et al. 2015, `\textit{Calibration of the Cherenkov Telescope Array}', these proceedings
\bibitem{simona} Toscano, S., et al. 2015, `\textit{Using muon rings for the calibration of the SST-1M prototype}', these proceedings
\bibitem{tom} Armstrong, T., et al. 2015, `\textit{Monte Carlo Studies of the GCT telescope for the Cherenkov Telescope Array}', these proceedings
\bibitem{cors} Heck, D., Knapp, J., Capdevielle, J.N., Schatz, G., and Thouw, T., 1998, `\textit{CORSIKA: a Monte Carlo code to simulate extensive air showers}', Report, FZKA, 6019    
\bibitem{telarr} Bernlohr, K., 2008, `\textit{Simulation of imaging atmospheric Cherenkov telescopes with CORSIKA and sim\_telarray}', APh, 30, 149
\end{thebibliography}
\end{document}